\documentclass{article}
\usepackage[utf8]{inputenc}
\usepackage[T1]{fontenc}
\usepackage{spconf,amsmath,graphicx,hyperref,multirow,bbding,pifont,etoolbox,fancyhdr}
\newcommand{\xmark}{\ding{55}}

\title{VoXtream: Full-Stream Text-to-Speech with Extremely Low Latency}
\name{Nikita Torgashov, Gustav Eje Henter, Gabriel Skantze}
\address{Department of Speech, Music and Hearing, KTH Royal Institute of Technology, Stockholm, Sweden}

\fancypagestyle{postprint}{%
  \fancyhf{} 
  \fancyfoot[C]{%
    \begin{minipage}{\textwidth}
    \begin{footnotesize}\begin{center}
    \textcopyright{} 2025 IEEE. Personal use of this material is permitted. Permission from IEEE must be obtained for all other uses, in any current or future media, including reprinting/republishing this material for advertising or promotional purposes, creating new collective works, for resale or redistribution to servers or lists, or reuse of any copyrighted component of this work in other works.
    \end{center}\end{footnotesize}
    \end{minipage}
  }
}

\begin{document}

\thispagestyle{postprint}

\makeatletter
\patchcmd{\thebibliography}
  {\advance\leftmargin\labelsep}
  {\setlength{\itemsep}{2pt plus 0.3pt}
   \setlength{\parsep}{1pt}
   \setlength{\parskip}{1pt}
   \advance\leftmargin\labelsep}
  {}
  {}
\makeatother

\maketitle

\begin{abstract}
We present VoXtream, a fully autoregressive, zero-shot streaming text-to-speech (TTS) system for real-time use that begins speaking from the first word. VoXtream directly maps incoming phonemes to audio tokens using a monotonic alignment scheme and a limited look-ahead that does not delay onset. Built around an incremental phoneme transformer, a temporal transformer predicting semantic and duration tokens, and a depth transformer producing acoustic tokens, VoXtream achieves, to our knowledge, the lowest initial delay among publicly available streaming TTS: 102\,ms on GPU. Despite being trained on a mid-scale 9k-hour corpus, it matches or surpasses larger baselines on several metrics, while delivering competitive quality in both output- and full-streaming settings. Demo and code are available at \url{https://herimor.github.io/voxtream}.

\end{abstract}


\begin{figure*}[t]
  \centering
  \includegraphics[width=430pt]{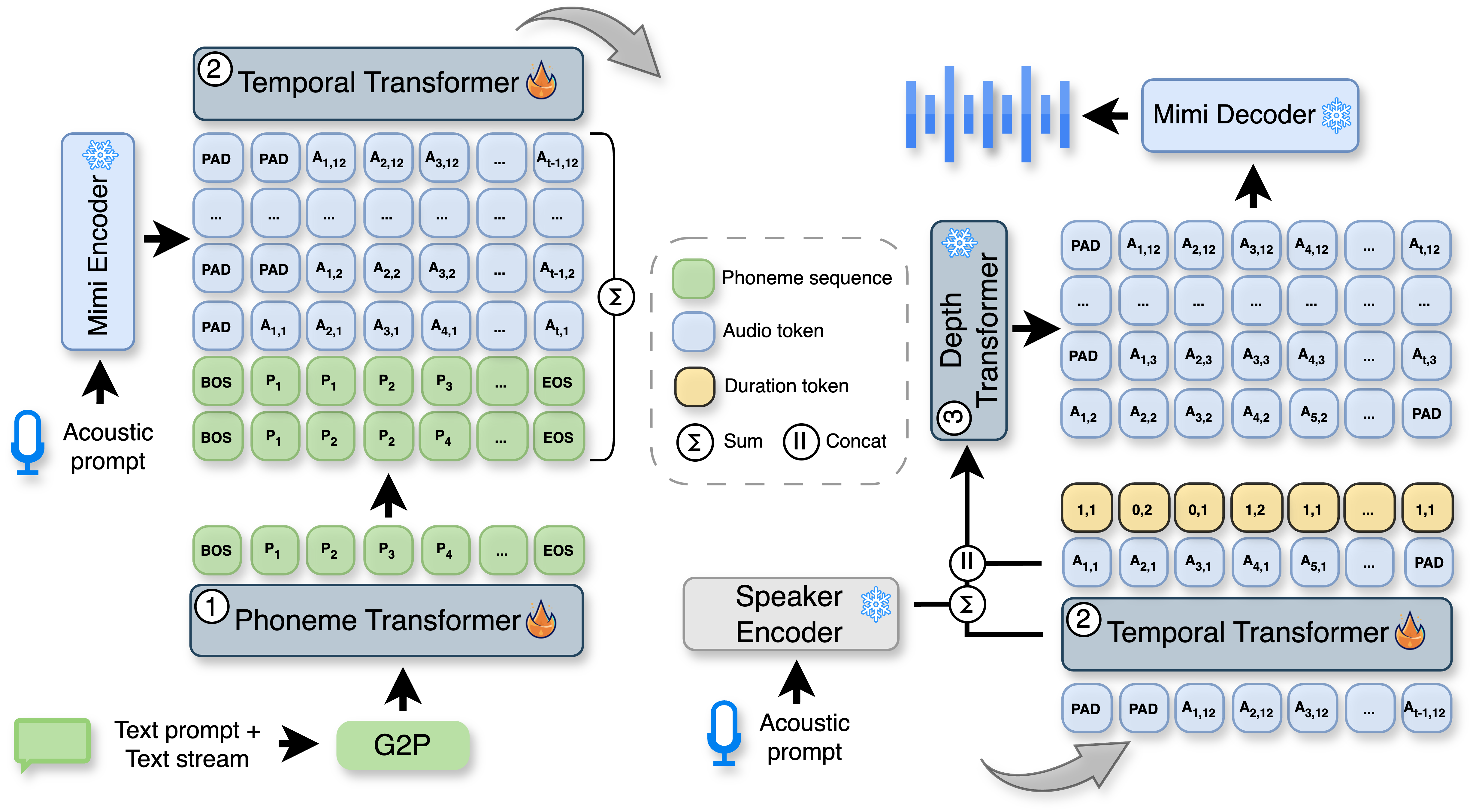}
  \caption{Overview of VoXtream, comprising an incremental Phoneme Transformer and Temporal and Depth Transformers.}
  \label{fig:arch}
\end{figure*}

\section{Introduction}

Recent advances in large language models (LLMs) and speech generation highlight the importance of low-latency, streaming text-to-speech (TTS) systems for real-time interaction. In applications such as voice assistants, simultaneous translation, and conversational AI, minimizing the delay between input and output -- especially first-packet latency -- is crucial for an engaging user experience. However, most existing TTS models either operate offline or rely on complex, multistage pipelines that limit responsiveness, require large input context, or introduce alignment challenges.

The rapid development of generative models, together with expanding training data, has produced multiple zero-shot TTS models with high naturalness and intelligibility on unseen speakers. Among autoregressive (AR) systems, two-stage models \cite{chen2025neural,du2024cosyvoice} first predict intermediate speech representations and then convert them to audio using a non-autoregressive decoder. Single-stage AR systems \cite{wang2025spark,ye2025llasa} directly map text to audio tokens with a single language model. While these models achieve high quality, they are fundamentally designed to process complete text segments before generating audio.

Several studies \cite{casanova24_interspeech,guo2025fireredtts} reduce latency via output streaming by generating speech in chunks as decoding proceeds. However, they do not address input-side latency and still require the entire text before speech generation begins.
Non-autoregressive (NAR) models based on language models \cite{wang2024maskgct}, factorized diffusion \cite{ju2024naturalspeech}, or flow matching \cite{le2023voicebox,mehta2024matcha,chen2025f5,zhu2025zipvoice} show strong offline performance, often surpassing AR models in speed and coherence, but their non-sequential architectures hinder streaming TTS.

Recent work in streaming TTS focuses on real-time synthesis through interleaved or chunked text-speech processing. SpeakStream~\cite{bai2025speakstream} employs a decoder-only transformer trained on interleaved text-speech sequences, achieving low first-packet latency but without exploring zero-shot capabilities. SyncSpeech~\cite{sheng2025syncspeech} introduces a temporal-masked transformer with one text-token look-ahead and a two-stage NAR decoder, which, however, requires accumulating tokens before audio decoding and thus adds delay.
IST-LM~\cite{yang2024interleaved} and CosyVoice2~\cite{du2024cosyvoice2} interleave text and speech at fixed ratios in an AR decoder to support full-stream operation, but both use the same NAR flow-matching decoder that operates on chunks, hurting first-packet latency.
Speak While You Think~\cite{dekel2024speak} distills from a non-streaming model with access to LLM embeddings and requires at least two words of look-ahead to maintain quality. LiveSpeech2~\cite{dang2024zero} breaks text into smaller segments and produces speech in a streaming manner but suffers from mismatches between speech generation and text fragments.

We introduce VoXtream, a fully autoregressive zero-shot TTS model that maps phoneme sequences to audio tokens with minimal delay. The architecture combines (1) an incremental Phoneme Transformer with limited phoneme look-ahead, (2) a Temporal Transformer (TT) that jointly predicts Mimi semantic and duration tokens, and (3) a Depth Transformer that generates acoustic tokens conditioned on outputs of TT and speaker embeddings. Unlike prior systems, the proposed method encodes text incrementally, making it possible to generate speech with a small look-ahead, as past text states are updated as more context is available. VoXtream begins outputting speech immediately after receiving the first word, achieving a first-packet latency as low as 102 ms on a modern GPU. Experiments conducted on SEED-TTS and LibriSpeech datasets show that the model generates highly intelligible speech, on par with state-of-the-art TTS models.

\section{Method}
\label{sec:method}

Figure \ref{fig:arch} presents an overview of the VoXtream architecture, which consists of three components:

\textbf{Phoneme Transformer}: The Phoneme Transformer (PT) is a decoder-only transformer that receives a phoneme sequence and encodes each phoneme with a corresponding embedding. The model operates incrementally, so its input grows with each new word from the text stream. For more natural prosody, PT is allowed to look ahead (LA) up to \textit{N} phoneme tokens. A key advantage is limited LA, where the model does not wait for \textit{N} phonemes before generating speech tokens but starts immediately after the first word. The minimal LA value is determined by the text-buffer size, and the maximal value is limited to 10 phonemes. We convert the input text stream to phonemes using g2pE\footnote{\url{https://github.com/Kyubyong/g2p}} at the word level.

\textbf{Temporal Transformer}: The Temporal Transformer (TT) is an autoregressive transformer conditioned on the audio tokens and the corresponding phoneme sequence. Audio tokens are extracted by the Mimi \cite{defossez2024moshi} codec at 12.5\,Hz. Due to the low frame rate, we assign up to two phonemes to each audio frame. Alignment between the acoustic prompt and the phoneme sequence is obtained using the Montreal Forced Aligner (MFA) \cite{mcauliffe2017montreal}. Following \cite{defossez2024moshi}, we apply an acoustic delay of one step for stability. The TT outputs the first codebook of Mimi (semantic tokens) and a duration token. Inspired by \cite{han2024vall}, we adopt monotonic phoneme alignment, but instead of predicting the specific phoneme, we predict a duration token that encodes a shift flag (stay/go, the first number in duration token), where ``stay" means that we continue generating of current phoneme in the next frame, or switch to the next phoneme otherwise, and the number of phonemes (1 or 2, the second number in duration token), corresponding to slower or faster pronunciation. Semantic and duration tokens are predicted by a single classification head and sampled from the joint distribution. The index of acoustic tokens corresponds to the temporal and codebook positions.

\textbf{Depth Transformer}: The Depth Transformer (DT) is an autoregressive transformer conditioned on the TT output embedding and the semantic token. As in \cite{defossez2024moshi}, DT generates audio tokens starting from the 2nd codebook of the Mimi codec (acoustic tokens). We also condition DT on a speaker embedding from ReDimNet \cite{yakovlev24_interspeech}. We use 12 Mimi codebooks as a latency-quality trade-off.

The Mimi decoder converts semantic and acoustic tokens for each frame into 80\,ms of speech in a streaming fashion. We train VoXtream by minimizing the negative log-likelihood of TT and DT outputs. An external aligner is used to obtain the initial alignment between the acoustic and text prompts.


\section{Experiments}
\label{sec:experiments}

\subsection{Experimental Setup}
\label{ssec:exp_setup}


\begin{table*}[t]
\caption{Evaluation results of zero-shot TTS. VoXtream-NS denotes the non-streaming version. Bold for the best result, underline for the second-best result (per section). Naturalness is measured as MUSHRA scores (0--100).}
\label{tab:eval}
\centering
\footnotesize
\setlength{\tabcolsep}{6pt}

\hspace*{-0cm}%
\resizebox{1.0\textwidth}{!}{
\begin{tabular}{@{}ll*{11}{c}@{}}
\hline \hline
  &
  \multirow{2}{*}{\textbf{Model}} &
  \textbf{Text} &
  \multirow{2}{*}{\textbf{\#Data(h)}} &
  \multirow{2}{*}{\textbf{\#Params}} &
  \multicolumn{3}{c}{\textbf{SEED \textit{test-en}}} &
  &
  \multicolumn{3}{c}{\textbf{LibriSpeech \textit{test-clean}}} &
  \textbf{Naturalness $\uparrow$} \\ 
  \cline{6-8}
  \cline{10-12}
  & & 
  \textbf{Token} &
  & &
  \textbf{WER (\%) $\downarrow$} &
  \textbf{SPK-SIM $\uparrow$} &
  \textbf{UTMOS $\uparrow$} &
  &
  \textbf{WER (\%) $\downarrow$} &
  \textbf{SPK-SIM $\uparrow$} &
  \textbf{UTMOS $\uparrow$} &
  \textbf{$\boldsymbol{\mu}$ $\pm$ 95\% CI} \\
\hline
& Human                & -     & -           & -    & 2.17             & 0.734             & 3.53      &        & 2.30             & 0.664             & 4.10            & 58.4 $\pm$ 2.5 \\
\hline
\multirow{6}*{\rotatebox{90}{\textbf{Large}}} &
  CosyVoice            & BPE   & 170k Multi. & 416M & 4.75             & \underline{0.635} & 3.88      &       & 3.75             & \underline{0.575} & 4.09             & -              \\
& Spark-TTS            & BPE   & 102k Multi. & 507M & 3.29             & 0.570             & 3.94      &       & \underline{3.02} & 0.513             & \underline{4.20} & -              \\
& Llasa-1B             & BPE   & 250k Multi. & 1000M & 3.18            & 0.578             & \underline{4.08} & & 3.18             & 0.490             & 4.19            & -              \\
& VoiceStar            & Phone & 65k  EN     & 840M & 2.91             & 0.605             & 3.92      &       & 3.92             & 0.509             & 4.10             & -              \\
& CosyVoice2           & BPE   & 167k Multi. & 618M & \underline{2.87} & \textbf{0.656}    & \textbf{4.18}   & & \textbf{2.97}    & \textbf{0.587}    & \textbf{4.23}    & -              \\
& FireRedTTS-1S        & BPE   & 500k Multi. & 550M & \textbf{2.66}    & 0.633             & 3.62      &       & 6.43             & 0.540             & 3.82             & -              \\
\hline
\multirow{3}*{\rotatebox{90}{\textbf{Mid}}} &
  VoiceCraft           & Phone & 9k   EN     & 830M & \underline{3.77} & \underline{0.515} & \underline{3.63} & & \underline{3.11} & \underline{0.444} & \underline{3.90} & 53.6 $\pm$ 2.5 \\
& XTTS-v2              & BPE   & 27k  Multi. & 470M & \textbf{3.64}    & 0.467             & 3.57             & & 3.90             & \underline{0.444}   & 3.72           & 53.8 $\pm$ 2.7 \\
& VoXtream-NS          & Phone & 9k   EN     & 441M & \textbf{3.64}    & \textbf{0.537}    & \textbf{3.89}    & & \textbf{2.99}    & \textbf{0.465}    & \textbf{4.07}    & 51.8 $\pm$ 2.6 \\
\hline
\multirow{4}*{\rotatebox{90}{\textbf{Stream}}} &
  CosyVoice2:\textbf{Out} & BPE   & 167k Multi. & 618M & \textbf{2.70}    & \textbf{0.662}    & \textbf{4.05}    & & \textbf{2.65}  & \textbf{0.592}    & \textbf{4.19}   & \textbf{60.6} $\pm$ 2.4 \\
& XTTS-v2:\textbf{Out}    & BPE   & 27k  Multi. & 470M & 3.99             & 0.480             & 3.59             & & 4.06           & 0.440             & 3.64            & 53.0 $\pm$ 2.7          \\
& VoXtream:\textbf{Out}   & Phone & 9k   EN     & 441M & 3.82             & \underline{0.529} & 3.88             & & \underline{3.09} & \underline{0.461} & \underline{4.08} & 53.4 $\pm$ 2.5       \\
& VoXtream:\textbf{Full}  & Phone & 9k   EN     & 441M & \underline{3.81} & \underline{0.529} & \underline{3.90} & & 3.15            & 0.458            & 4.07            & 51.9 $\pm$ 2.6          \\
\hline \hline
\end{tabular}%
}
\vspace{-1ex}
\end{table*}

\begin{table}[h!]
\caption{Full-stream capabilities on the LibriSpeech \textit{long} set. Pref. denotes naturalness preference.}
\label{tab:eval_stream}
\centering
\footnotesize
\setlength{\tabcolsep}{2pt} 

\hspace*{-0.0cm}
\resizebox{0.45\textwidth}{!}{
\begin{tabular}{lcccc}
\hline \hline
  \textbf{Model} &
  \textbf{WER (\%) $\downarrow$} &
  \textbf{SPK-SIM $\uparrow$} &
  \textbf{UTMOS $\uparrow$} &
  \textbf{Pref.(\%) $\uparrow$} \\
\hline
Human                    & 1.97          & 0.784          & 4.16          & -           \\
\hline
CosyVoice2:\textbf{Full} & 6.11          & \textbf{0.685} & 4.19          & 31          \\
VoXtream:\textbf{Full}   & \textbf{3.24} & 0.564          & \textbf{4.23} & \textbf{57} \\ 
\hline \hline
\end{tabular}
}
\end{table}

\textbf{Dataset}: Our training corpus is based on the Emilia \cite{he2024emilia} and HiFiTTS-2 \cite{langman25_interspeech} datasets, covering both spontaneous and read styles. We selected 4.5k hours from each, totaling 9k hours, to be comparable to \cite{peng2024voicecraft} in dataset size. For Emilia, we applied additional diarization to remove multi-speaker utterances and discarded utterances with invalid automatic transcripts. We also used NISQA \cite{mittag21_interspeech} to remove low-quality utterances. We used the CMU dictionary to map text into phonemes and aligned them to audio with MFA \cite{mcauliffe2017montreal}. Speech tokenization used the streaming Mimi \cite{defossez2024moshi} codec at 24\,kHz.

\textbf{Model}: We used a Llama-style \cite{dubey2024llama} transformer for all modules. The Temporal Transformer has 12 layers, 16 attention heads, an embedding dimension of 1024, and a feed-forward dimension of 4096, similar to \cite{chen2025neural}. The Phoneme and Depth transformers share a similar design with minor differences: the Phoneme Transformer has 6 layers and 8 heads; the Depth Transformer has 4 layers, 8 heads, and a feed-forward dimension of 8192. We borrowed the Depth Transformer from the CSM model\footnote{\url{https://github.com/SesameAILabs/csm}} (CSM-DT) trained on a large-scale dataset and kept its weights frozen during training. As a speaker encoder, we used ReDimNet \cite{yakovlev24_interspeech} (SPK-ENC) trained on 100k+ identities.

Models were trained on two NVIDIA A100-80GB GPUs with a batch size of 128 per GPU for 9 epochs. We used fixed 20\,s audio chunks and their corresponding phoneme sequences as input. Because most utterances are shorter, we concatenated multiple utterances within the same speaker. We used AdamW \cite{loshchilov2017decoupled}, warming up the learning rate for the first epoch to a peak of $5 \times 10^{-4}$.


\textbf{Baseline Models}: We chose multiple AR models with publicly available weights for comparison. For the non-streaming scenario we selected CosyVoice \cite{du2024cosyvoice}, Spark-TTS \cite{wang2025spark}, Llasa \cite{ye2025llasa}, VoiceStar \cite{peng2025voicestar}, VoiceCraft \cite{peng2024voicecraft}, XTTS \cite{casanova24_interspeech}, CosyVoice2 \cite{du2024cosyvoice2}, and FireRedTTS-1S \cite{guo2025fireredtts}. We split them into Large/Mid groups by training data size for fair comparison.

For streaming evaluation, we selected XTTS \cite{casanova24_interspeech} and CosyVoice2 \cite{du2024cosyvoice2}. FireRedTTS-1S \cite{guo2025fireredtts} was omitted because the authors did not release the streaming generation code. We also noticed that CosyVoice2 re-synthesizes the prompt in the full-stream scenario when the target text is shorter than the prompt, inflating WER, so we do not report those numbers. Instead, we compare VoXtream to CosyVoice2 in the full-stream scenario on LibriSpeech \textit{long}, where this issue does not occur.

\subsection{Evaluation}
\label{ssec:evaluation}

We evaluated VoXtream on three test sets. The first is LibriSpeech \cite{panayotov2015librispeech} \textit{test-clean}. Following \cite{chen2025neural,yang2024interleaved}, we extracted a 2.2\,h subset and evaluated the continuation task. The second is SEED-TTS \cite{anastassiou2024seed} \textit{test-en} for a cross-sentence task. The third set targets long-sequence robustness in the full-stream setting: we selected utterances longer than 10\,s from LibriSpeech \textit{test-clean}, yielding a 2.5\,h subset with an average length of 15\,s, denoted LibriSpeech \textit{long}.

We used three reproducible model-based metrics. For intelligibility, we report WER between the transcription of synthesized speech and the input text. For SEED-TTS \textit{test-en} we used Whisper-large-v3 \cite{radford2023robust} and followed metric calculation from the official SEED test repository. For LibriSpeech \textit{test-clean} we used a HuBERT-based ASR \cite{hsu2021hubert} and prepended the audio prompt to the generated continuation for WER, as in \cite{yang2024interleaved}. For speaker similarity, we computed cosine similarity (SPK-SIM) between embeddings from a WavLM-based \cite{chen2022wavlm} ECAPA-TDNN \cite{desplanques20_interspeech} for the prompt and synthesized speech. For quality, we used the UTMOS \cite{saeki22c_interspeech} MOS predictor.

To evaluate naturalness, we conducted two user studies on the Prolific platform. Each study included multiple attention checks, and participants were compensated at an average rate of 9 GBP per hour. For a fair comparison, we sampled 100 unique sentences on which all systems achieved 0\% WER. The first study focused on short-form TTS. We recruited 40 native listeners and asked them to rate the naturalness of each system on a 0--100 scale, following a MUSHRA-like protocol. The second study focused on full-stream evaluation. Here, 30 native listeners were asked to select their preferred system or indicate ``no preference".

In the full-stream evaluation, text was provided word by word to simulate streaming input from an LLM. We also measured first-packet latency (FPL), the time to the first speech frame, and the real-time factor (RTF), defined as the ratio of generated speech duration to wall-clock generation time.



\begin{table}[t!]
\caption{Performance in FP16 on an A100 GPU. \textbf{TC} denotes \texttt{torch.compile} and \textbf{DS} denotes \texttt{DeepSpeed}. XTTS-v2 metrics are for output streaming; other models are full-stream.}
\label{tab:rtf}
\centering
\footnotesize
\hspace*{-0.0cm}

\resizebox{0.3\textwidth}{!}{
\begin{tabular}{lcc}
\hline \hline
  \textbf{Model} &
  \textbf{FPL(ms) $\downarrow$} &
  \textbf{RTF} $\downarrow$ \\
\hline
CosyVoice2       & 1643          & 0.85          \\
XTTS-v2          & 295           & 0.37          \\
XTTS-v2:DS       & 196           & 0.26          \\
VoXtream         & 171           & 1.00          \\
VoXtream:TC      & \textbf{102}  & \textbf{0.17} \\
\hline \hline
\end{tabular}
}
\end{table}

\begin{table}[t!]
\caption{Ablation study on foundation-model components on SEED \textit{test-en}.}
\label{tab:ablation}
\centering
\footnotesize
\setlength{\tabcolsep}{2pt}
\hspace*{-0.0cm}

\resizebox{0.42\textwidth}{!}{
\begin{tabular}{ccccc}
\hline \hline
  \textbf{CSM-DT} &
  \textbf{SPK-ENC} &
  \textbf{WER (\%) $\downarrow$} &
  \textbf{SPK-SIM $\uparrow$} &
  \textbf{UTMOS $\uparrow$} \\
\hline
\xmark     & \xmark     & \textbf{3.53}    & 0.471             & 3.39             \\ 
\Checkmark & \xmark     & 3.70             & 0.504             & \textbf{3.90}    \\ 
\xmark     & \Checkmark & 3.65             & \textbf{0.558}    & 3.39             \\ 
\Checkmark & \Checkmark & \underline{3.64} & \underline{0.537} & \underline{3.89} \\ 
\hline \hline
\end{tabular}
}
\end{table}

\section{Results and Discussion}
\label{sec:results}

Table \ref{tab:eval} presents the results for VoXtream and baselines. With comparable training data (Mid-scale), our model achieves the best SPK-SIM and UTMOS metrics and is comparable in terms of WER. We report two variants: a non-streaming model with effectively infinite look-ahead and a streaming model with limited phoneme look-ahead. VoXtream maintains high quality even with limited future context, introducing only slight degradations in WER and SPK-SIM relative to the non-streaming variant.


While less accurate than models trained on Large-scale datasets, VoXtream approaches Spark-TTS and CosyVoice despite much less data, while also providing streaming. Moreover, VoXtream attains the second-best WER on LibriSpeech \textit{test-clean} among all systems, confirming high intelligibility.

In short-form streaming, VoXtream outperforms XTTS and is second only to CosyVoice2 in output streaming. In full-stream, WER is slightly higher because the input arrives word by word; SPK-SIM and UTMOS are on par with output streaming, where the full text is known before generation. Subjective evaluation also shows that our streaming model is comparable to non-streaming VoiceCraft and XTTS models.

Table \ref{tab:eval_stream} shows the results of full-stream TTS on LibriSpeech \textit{long} set. VoXtream delivers much lower WER compared to CosyVoice2, and naturalness favors our model in subjective evaluation ($p<5e^{-10}$). CosyVoice2 has a higher speaker similarity due to its NAR flow-matching decoder, but this comes with a high FPL.

A key contribution is the extremely low initial latency (Table \ref{tab:rtf}). VoXtream has the lowest FPL among publicly available streaming TTS models and, with \texttt{torch.compile}, reaches 102\,ms. VoXtream runs in real time on the GPU without extra acceleration and, with \texttt{torch.compile}, runs more than $5\times$ faster than real time, achieving the lowest RTF among streaming competitors.

We also conducted an ablation study to assess foundation-model components (Table \ref{tab:ablation}). As a baseline, we removed SPK-ENC and trained DT from scratch. We then added either SPK-ENC or CSM-DT, and finally both. CSM-DT substantially improves quality and speaker similarity via knowledge transfer. SPK-ENC increases SPK-SIM by 19\% in zero-shot TTS when DT is unfrozen and by 6\% even with a frozen DT. For intelligibility, the baseline achieves the best WER, highlighting the efficiency of the proposed method; the slight increase in WER of the final system is not significant.

\section{Conclusions and future work}

We introduced VoXtream, a full-stream zero-shot TTS model. The method delivers real-time synthesis with ultra-low initial latency, achieving the best performance among publicly available streaming models. The use of foundation models helps VoXtream to match or exceed the quality of larger (in terms of parameters or training data) non-streaming and streaming baselines on several benchmarks, and provides strong performance in both output- and full-stream settings with only minor degradations relative to its non-streaming variant.

In future work, we plan to explore scaling the training data, explicit speaking-rate control, and long-form streaming speech synthesis.

\section{Acknowledgements}
\label{sec:ackn}

This work was partially supported by the Wallenberg AI, Autonomous Systems and Software Program (WASP) funded by the Knut and Alice Wallenberg Foundation (KAW) and by the Industrial Strategic Technology Development Program (grant no.\ 20023495) funded by MOTIE, South Korea. Computations were enabled by the supercomputing resource Berzelius provided by KAW and NSC at Linköping University.

\bibliographystyle{IEEEbib}
\bibliography{refs}

\end{document}